\documentclass[12pt]{article}
\usepackage{amsxtra,amssymb,amsthm,amsmath,latexsym}
\usepackage{float}
\usepackage{graphicx}
\restylefloat{table}

\theoremstyle{plain}

\def\oH{{\overset{\circ}{H}}}
\def\oH1{{\overset{\circ}{H}\kern-.02in{}^1}}

\def\bee{\begin{equation*}}
\def\eee{\end{equation*}}
\def\be{\begin{equation}}
\def\ee{\end{equation}}

\begin{document}

\title{Representation of big data by dimension reduction}

\author{A.G.Ramm, C. Van\\
 Department  of Mathematics, Kansas State University, \\
 Manhattan, KS 66506, USA\\
ramm@math.ksu.edu; congvan@math.ksu.edu\\
}

\date{}
\maketitle\thispagestyle{empty}

\abstract{Suppose the data consist of a set $S$ of points $x_j, 1 \leq j \leq J$, distributed in a bounded domain $D \subset R^N$, where $N$ and $J$ are large numbers. In this paper an algorithm is proposed for checking whether there exists a manifold $\mathbb{M}$ of low dimension near which many of the points of $S$ lie and finding such $\mathbb{M}$ if it exists. There are many dimension reduction algorithms, both linear and non-linear. Our algorithm is simple to implement and has some advantages compared with the known algorithms. If there is a manifold of low dimension near which most of the data points lie, the proposed algorithm will find it. Some numerical results are presented illustrating the algorithm and analyzing its performance compared to the classical PCA (principal component analysis) and Isomap.
}






\section{Introduction} \label{Introduction}

\noindent There is a large literature on dimension reduction. Without trying to describe the published results, we refer the reader to the reviews \cite{Fodor} and \cite{Maaten} and references therein. Our algorithm is simple, it is easy to implement, and it has some advantages over the known algorithms. We compare its performance with PCA and Isomap algorithms. The main point in this short paper is the new algorithm for computing a manifold $\mathbb{M}$ of low dimension in a neighborhood of which most of the data points lie, or finding out that there is no such manifold.

\noindent In section 2, we describe the details of the proposed algorithm. In section 3, we analyze the performance of the algorithm and compare its performance with PCA and Isomap algorithms' performances. In section 4, we show some numerical results.

\section{Description of the algorithm} \label{Description of the algorithm}

\noindent Let $S$ be the set of the data points $x_j, 1 \leq j \leq J$, $x_j \in D \subset R^N$, where $D$ is a unit cube and $J$ and $N$ are very large. Divide the unit cube $D$
into a grid with step size $0 < r \leq 1$. Let $a = \frac{1}{r}$ be the number of intervals on each side of the unit cube $D$. Let $V$ be the upper
limit of the total volume of the small cubes $C_m$, $L$ be the lower limit of the number of the data points near the manifold $\mathbb{M}$, and $p$ be the lower limit of the
number of the data points in a small cube $C_m$ with the side $r$. By a manifold in this paper, a union of piecewise-smooth manifolds is meant. A smooth manifold is defined to be a smooth lower-dimensional domain in $\mathbb{R}^N$, for example, a line in $\mathbb{R}^3$, or a plane in $\mathbb{R}^3$. The one-dimensional manifold that we construct is a union of piecewise-linear curves, a linear curve is a line.

\noindent Let us describe the algorithm, which is a development of an idea in \cite{Ramm}. The steps of this algorithm are:
\begin{enumerate}
\item[1.] Take $a = a_1$, and $a_1 = 2$.
\item[2.] Take $M = a^N$, where $M$ is the number of the small cubes with the side $r$. Denote these small cubes by $C_m, 1 \leq m \leq M$. Scan the data set
$S$ by moving the small cube with the step size $r$ and calculating the number of points in this cube for each position of this cube in $D$. Each of the points of $S$ lie in some
cube $C_m$. Let $\mu_m$ be the number of the points of $S$ in $C_m$. Neglect
 the small cubes with $\mu_m < p$, where $p$ is the chosen lower limit of the number of the data points in a small cube $C_m$. One chooses $p \ll |S|$, for example, $p=0.005 |S|$, where $|S|$ is the number of the data points in $S$. Let $V_t$ be the total volume of the cubes that we keep.
\item[3.] If $V_t = 0$, we conclude that there is no manifold found.
\item[4.] If $V_t > V$, we set $a = a_2 = 2a_1$. If $a_2 > \sqrt[N]{|S|}$, we conclude that there is no manifold found. Otherwise, repeat steps 2 and 3.
\item[5.] If $V_t < V$, then denote by $C_k, 1 \leq k \leq K \ll M$, the small cubes that we keep ($\mu_k \geq p$). Denote the centers of $C_k$ by $c_k$ and let $\mathcal{C} := \cup_{k=1}^K C_k$.
\item[6.] If the total number of the points in $\mathcal{C}$ is less than $L$, where $L$ is the chosen lower limit of the number of the data points near the manifold $\mathbb{M}$, then we conclude that there is no manifold found.
\item[7.] Otherwise, we have a set of small cubes $C_k, 1 \leq k \leq K \ll M$ with the side $r$, which has total volume less than $V$ and the
number of the data points $\geq L$.
\end{enumerate}

\noindent Given the set of small cubes $C_k, 1 \leq k \leq K$, one can build the sets $L_s$ of dimension $s \ll N$, in a neighborhood of which maximal
 amount of points $x_j \in S$ lie. For example, to build $L_1$, one can start with the point $y_1 := c_1$ where $C_1$ is the small cube closest to the origin, and join $y_1$ with $y_2 = c_2$ where $C_2$ is the small cube closest to $C_1$. Continuing joining $K$ small cubes, one gets a one-dimensional piecewise-linear manifold $L_1$. To build $L_2$, one considers the triangles $T_k$ with vertices $y_k, y_{k + 1}, y_{k + 2}$, $1 \leq k \leq K - 2$. 
   The union of $T_k$ forms a two-dimensional manifold $L_2$. Similarly, one can build $s$-dimensional manifold $L_s$ from the set of $C_k$.\\

\noindent While building  manifolds $L_s$, one can construct several such manifolds because the closest to $C_{i-1}$ cube $C_i$ may be non-unique.
 However, each of the constructed manifolds contains the small cubes $C_k$, which have totally at least $L$ data points. For example, for
  $L = 0.9|S|$ each of these manifolds contains at least $0.9|S|$ data points. Of course, there could be a manifold containing $0.99 |S|$ and
   another one containing $0.9|S|$. But if the goal is to include as many data points as possible, the experimenter can increase $L$, for example,
    to $0.95|S|$.\\

\noindent \textbf{Choice of $V, L,$ and $p$}\\

\noindent The idea of the algorithm is to find a region containing most of the data points ( $\geq L$ ) that has small volume ( $\leq V$ ) by
neglecting the small cubes $C_k$ that has $< p$ data points.\\

\noindent Depending on how sparse the data set is, one can choose an appropriate $L$. If the data set has a small number of outliers ( an outlier is
an observation point that is distant from other observations), one can choose $L = 0.9 |S|$, as in the first numerical experiment in Section 3. If the
data set is sparse, one can choose $L = 0.8 |S|$, as in the second numerical experiment in Section 3. In general, when the experimenter does not know
how sparse the data are, one can try $L$ to be $0.95 |S|$, $0.9 |S|$ or $0.8|S|$.\\

\noindent One should not choose $V$ too big, for example $\geq 0.5$, because it does not make sense to have a manifold with big volume. One should not
 choose $V$ too small because then either one cannot find a manifold or the number of small cubes $C_k$ is the same as the number of data points. So,
 the authors suggest $V$ to be between $0.3$ and $0.4$ of the volume of the original unit cube where the data points are.\\

\noindent The value $p$ is used to decide whether one should neglect the small cubes $C_k$. So, it is recommended to choose  $p<<|S|$. Numerical
experiments show that $p = 0.005|S|$ works well. The experimenter can also try smaller values of $p$ but then more computation time is needed.

\noindent

\section{Performance analysis}\label{Performance analysis}

\noindent In the algorithm one doubles $a$ each time and $2 \leq a \leq \sqrt[N]{|S|}$. So the main step runs at most
$\log \sqrt[N]{|S|}$ or $\frac{1}{N}\log |S|$ times. For each $a$ one calculates the number of points of the data set $S$ in each of $M_a$ small cubes.
 The total computational time is $|S| \, M_a \, \frac{1}{N}\log |S|$. In $N$-dimensional space, calculating the distance between two points takes $N$
 operations. Thus, the asymptotic speed of the algorithm is $O\left( N \, |S| \, M_a \, \frac{1}{N}\log |S| \right)$ or
  $O\left( |S| \, M_a \,\log |S| \right)$. Since $M_a \leq |S|$, in the worst case, the asymptotic speed is $O(|S|^2 \, \log |S|)$.

\noindent If one compares this algorithm with the principal component analysis (PCA) algorithm, which has asymptotic speed $O(|S|^3)$, one
sees that our algorithm is much faster than the PCA algorithm. PCA theory is presented in \cite{J}.

\noindent In paper \cite{Cayton} the fastest algorithm for finding a manifold of lower dimension containing many points of S, called Isomap,
 has the asymptotic speed of $O(s \, l \, |S| \log |S|)$, where $s$ is the assumed dimension of the manifold and $l$ is the number of landmarks.
  However, in the Isomap algorithm one has to use a priori assumed dimension of the manifold, which is not known a priori, while our algorithm finds
   the manifold and its dimension. Also, in the Isomap algorithm one has to specify the landmarks, which can be arbitrarily located in the domain $D$.
    Our algorithm is simpler to implement than the Isomap algorithm.

\noindent For large $|S|$, one can make our algorithm faster by putting the upper limit for $a$. For example, instead of 
$1 \leq a \leq \sqrt[N]{|S|}$, one can require $1 \leq a \leq \sqrt[2N]{|S|}$.

\section{Numerical results}
In the following numerical experiments, we use different data sets from paper \cite{Chang} and \cite{Jane}. We apply our proposed algorithm above to each data set and get the results as shown in the pictures. The first picture of each data set shows the original data points (as blue points) and the found small cubes (as red points). The second picture of each data set shows the found manifold.
\begin{itemize}
\item[1.] Consider the data set from paper \cite{Chang}. This is a 2-dimensional set containing $|S| = 308$ points in a 2-dimensional Cartesian
coordinate system. Below we take the integer value for $p$ nearest to $p$ and larger than $p$.

\begin{figure}[H]
  \centering
      \includegraphics[width=\textwidth/2]{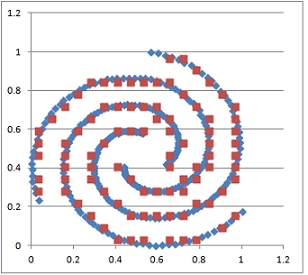}
  \caption{a data set from paper \cite{Chang} with $|S| = 308$ and $p = 0.005|S| = 1.56$}
\end{figure}

\begin{figure}[H]
  \centering
      \includegraphics[width=\textwidth/2]{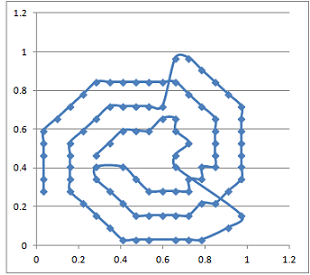}
  \caption{a data set from paper \cite{Chang} with $|S| = 308$ and $p = 0.005|S| = 1.56$}
\end{figure}

\noindent Run the algorithm with $V = 0.5$ (i.e. the final set of $C_k$ has to have total volume less than $0.5$),
 $L = 0.9 \times |S|$ (the manifold has to contains at least $90\%$ of the initial data points), and $p = 0.005|S| = 1.56$ (we take $p=2$ and
 neglect the
 small cubes that have  $\le 1$ point). One finds the manifold $L_1$ when $a = 16$, $r = 1/16$, the number of small cubes
  $C_k$ is $M = 89$, which has total volume of $0.35$ and contains $0.92|S|$ data points.\\

\item[2.] Consider another data set from paper \cite{Chang}. This set has $|S| = 296$ data points.

\begin{figure}[H]
  \centering
      \includegraphics[width=\textwidth/2]{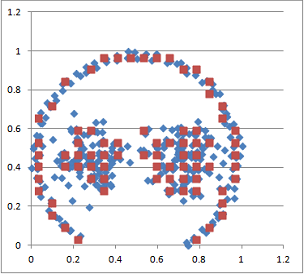}
  \caption{another data set from paper \cite{Chang} with $|S| = 296$ and $p = 0.005|S| = 1.5$}
\end{figure}

\begin{figure}[H]
  \centering
      \includegraphics[width=\textwidth/2]{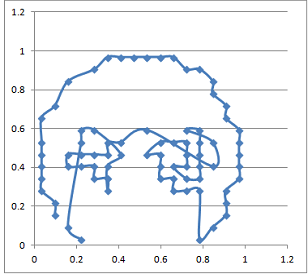}
  \caption{another data set from paper \cite{Chang} with $|S| = 296$ and $p = 0.005|S| = 1.5$}
\end{figure}

\noindent Run the algorithm with $V = 0.5$ (i.e. the final set of $C_k$ has to have total volume less than $0.5$), $L = 0.8 \times |S|$
(since some part of the data is uniformly distributed), and $p = 0.005|S| = 1.5$ (we take $p=2$ and  neglect the small cubes that have
 $\le 1$ point). One finds the manifold $L_1$ when $a = 16$, $r = 1/16$, the number of small cubes $C_k$ is $M = 76$, which has total volume of
 $0.3$ and contains $0.83|S|$ data points.\\

\item[3.] Use the data set from paper \cite{Chang} as in the second experiment. In this experiment, $p = 4$ was used. This is to show that one
can try different values of $p$ (besides the suggested one $p = 0.005|S|$) and may find a manifold with a smaller number of small cubes.

\begin{figure}[H]
  \centering
      \includegraphics[width=\textwidth/2]{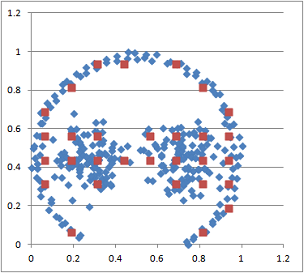}
  \caption{another data set from paper \cite{Chang}  with $|S| = 296$ and $p = 4$}
\end{figure}

\begin{figure}[H]
  \centering
      \includegraphics[width=\textwidth/2]{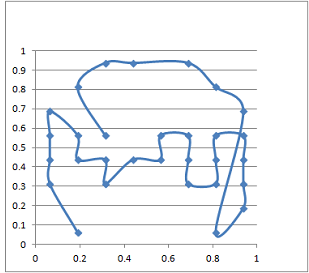}
  \caption{another data set from paper \cite{Chang}  with $|S| = 296$ and $p = 4$}
\end{figure}

\noindent Choose $p = 4$, $V = 0.5$ and $L = 0.8 \times |S|$. One finds the manifold $L_1$ when $a = 8$, $r = 1/8$, the number of small cubes
$C_k$ is $M = 30$ ( less than $76$ in experiment 2), which has total volume of $0.47$ and the total number of points  $0.89|S|$.
 The conclusion is:  one can try different values for $p$ and can have different manifolds.

\item[4.] Consider a data set from paper \cite{Fu} (Figure 7).

\begin{figure}[H]
  \centering
	\includegraphics{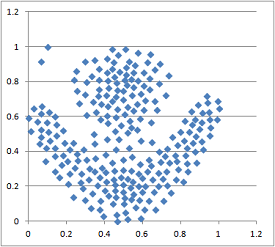}
  \caption{a data set from paper \cite{Fu} with $|S| = 235$}
\end{figure}

\noindent Run the algorithm with $V = 0.5$, $L = 0.8 \times |S|$ and let $p$ run from $1$ to $10$. For any $p$, one cannot find a set of
$C_k$ which has total volume $\le V$ and contains at least $0.8|S|$ data points. This data set does not have a manifold
of lower dimension in a neighborhood of which most of the data points lie.\\\\

\item[5.] Consider a data set from paper \cite{Jane}. This set has $|S| = 373$ data points.

\begin{figure}[H]
  \centering
      \includegraphics[width=\textwidth/2]{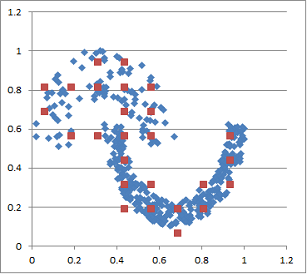}
  \caption{a data set from paper \cite{Jane} with $|S| = 373$ and $p = 4$}
\end{figure}

\begin{figure}[H]
  \centering
      \includegraphics[width=\textwidth/2]{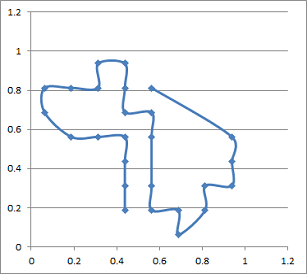}
  \caption{a data set from paper \cite{Jane} with $|S| = 373$ and $p = 4$}
\end{figure}

\noindent Run the algorithm with $V = 0.5$ (i.e. the final set of $C_k$ has to have total volume less than $0.5$), $L = 0.8 \times |S|$
(since some part of the data is uniformly distributed), and $p = 4$ (we neglect the small cubes that have
 $\le 3$ points). One finds the manifold $L_1$ when $a = 8$, $r = 1/8$, the number of small cubes $C_k$ is $M = 26$.\\
\newpage
\item[6.] Use the data set from paper \cite{Jane} as in the fifth experiment. In this experiment, $p = 2$ was used. This is to show that one
can try different values of $p$ (besides the suggested one $p = 4$) and may find a different manifold.

\begin{figure}[H]
  \centering
      \includegraphics[width=\textwidth/2]{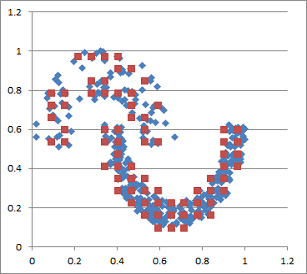}
  \caption{a data set from paper \cite{Jane}  with $|S| = 373$ and $p = 2$}
\end{figure}

\begin{figure}[H]
  \centering
      \includegraphics[width=\textwidth/2]{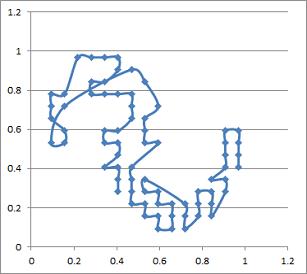}
  \caption{a data set from paper \cite{Jane}  with $|S| = 373$ and $p = 2$}
\end{figure}

\noindent Choose $p = 2$, $V = 0.5$ and $L = 0.8 \times |S|$. One finds the manifold $L_1$ when $a = 16$, $r = 1/16$, the number of small cubes
$C_k$ is $M = 72$. The conclusion is:  one can try different values for $p$ and can have different manifolds.
\end{itemize}
$\\$
\section{Conclusion}\label{Conclusion}

We conclude that our proposed algorithm can compute a manifold $\mathbb{M}$ of low dimension in a neighborhood of which most of the data points lie, or find out that there is no such manifold. Our algorithm is simple, it is easy to implement it, and it has some advantages over the known algorithms (PCA and Isomap).

\end{document}